# Fluid Flow Complexity in Fracture Networks: Analysis with Graph Theory and LBM

Ghaffari, H.O.
*Department of Civil Engineering and Lassonde Institute, University of Toronto, Toronto, Canada*

Nasseri, M.H.B.
*Department of Civil Engineering and Lassonde Institute, University of Toronto, Toronto, Canada*

Young, R.P.
*Department of Civil Engineering and Lassonde Institute, University of Toronto, Toronto, Canada*

**ABSTRACT:** Through this research, embedded synthetic fracture networks in rock masses are studied. To analysis the fluid flow complexity in fracture networks with respect to the variation of connectivity patterns, two different approaches are employed, namely, the Lattice Boltzmann method and graph theory. The Lattice Boltzmann method is used to show the sensitivity of the permeability and fluid velocity distribution to synthetic fracture networks' connectivity patterns. Furthermore, the fracture networks are mapped into the graphs, and the characteristics of these graphs are compared to the main spatial fracture networks. Among different characteristics of networks, we distinguish the modularity of networks and sub-graphs distributions. We map the flow regimes into the proper regions of the network's modularity space. Also, for each type of fluid regime, corresponding motifs shapes are scaled. Implemented power law distributions of fracture length in spatial fracture networks yielded the same node's degree distribution in transformed networks. Two general spatial networks are considered: random networks and networks with "hubness" properties mimicking a spatial damage zone (both with power law distribution of fracture length). In the first case, the fractures are embedded in uniformly distributed fracture sets; the second case covers spatial fracture zones. We prove numerically that the abnormal change (transition) in permeability is controlled by the hub growth rate. Also, comparing LBM results with the characteristic mean length of transformed networks' links shows a reverse relationship between the aforementioned parameters. In addition, the abnormalities in advection through nodes are presented.

## 1. INTRODUCTION

Fracture networks embedded in different structures are subject to the varied research areas. However, the main relevance of fracture networks is the propagation of fractures (cracks) in brittle materials. The complex form of linked cracks has inspired scientists to develop several theories of fracture network formation's general mechanisms. One of these theoretical branches is associated with single fracture mechanics. In complement with this approach, the interaction of fractures yields modified stress field. Then, the concentration of stress or variation of strain satisfies yielding criteria [1]. Inducing disorder into the system (say intact rock) gives the fracture's path of propagation. This branch of fracture propagation includes a wide range of methods: such as, linear elastic fracture methods; effective medium theory; finite or boundary element methods to tackle the complex geometries; molecular dynamics-based methods (i.e., distinct element methods) to give detailed insight into the system; lattice-based methods associated with random fuse models or random beam methods; and, finally, the methods based on statistical field theory like interface or string growth methods [2-5].

The second approach corresponds to the collective observations from controllable laboratory tests and field data. The obtained patterns thus are analyzed using pattern recognition techniques. Eventually, the aim is to organize simple rules to satisfy the same characteristics (features or attributes) for the fracture system. Such simplistic methods have originated from associated with statistical physics. To be specific, the most well-known employed methods fall under the second approach, applying statistical techniques to fracture networks. It satisfies general characteristics of fracture networks like distribution law over density, length, directionality, distance, and fractal dimension— as well as characteristics of flow propagation like permeability [6-9].

The first steps in analyzing the fractures' structural configurations included the fractality of the fracture systems. Having a fractal dimension with a small variation indicates that the system has a universality property. However, there is not such universal dimensionality in fracture networks. Researchers have recognized two fractal dimensions for the roughness (non-directional) of a single fracture [10]. During the past decade, developments in graph theory [11-15] have opened a new chapter in the study of large interwoven systems (complex systems). This new perspective tackles another facet of the complexities embedded in fractured materials, called structural or topological complexity. In contrast, statistical methods investigate uncertainties such as probability or, in more advanced forms, other uncertainties approaches (like fuzzy set theory, rough set theory or evidence theories). The relevant topics in this field might be summarized as follows: 1) transformation of spatial networks in graph forms; 2) information propagation (fluid flow, energy or waves) through networks; 3) deformation of fractures due to mechanical or hydro-mechanical forces and, more generally, the reaction of a disordered system (like a fracture system) to external forces.

Our focus in this study is to generate different configurations of spatial fracture networks. In this pursuit, we try to use the simplest algorithms to fracture network generation. Then, we transform spatial fracture networks into graph forms where we ignore the spatial distribution of fractures. The advantage of these transformations is that it links the regular fracture networks to modern graph theory. The fracture zones as a high density of fractures also as the abnormal emergence of fractures are another part of our research. We present a simple algorithm that gives the directionality and effective radius of the hubs, while taking into account the growth of the hubs under a background growth of random joints. Advection of information (here fluid flow) through the generated spatial network and transformed networks is the main part of this study.

To analyze fluid flow, three different methods were employed. We use lattice Boltzmann (LBM) and finite element methods (FEM) to obtain the fluid flow patterns, velocity, pressure distribution and permeability for the spatial fracture networks. Additionally, an advection- based network equation is used for transformed networks. The organization of the paper is as follows. The first section presents a simple algorithm to generate fracture networks and show how they are converted into the graphs. In this section, we also introduce some basic characteristics of graphs. The next section summarizes three methods— LBM, FEM and advection-based networks— to model fluid flow (laminar) in fracture networks. The main part of our study will be covered in section 4, where we present the results and discuss the accuracy of the employed methods. Finally, the summary and conclusion of the present work is presented.

## 2. FRACTURE NETWORKS AND GRAPH THEORY

In this part, a simple algorithm is introduced to cover the main characteristics of fracture networks. The transformation of constructed fracture networks into graph forms, and the distinguishing characteristics of the graphs, are demonstrated next.

### *2.1. Random Fracture Networks*

Several algorithms have been proposed to generate and cover the main statistical properties of natural fracture networks (based on the second approach presented in our introduction). The simplest algorithms in 2D consider the distribution of fracture length, dip direction of joint sets and joint spacing [7-9]. New generations of fracture networks algorithms (or Discrete Fracture Networks: DFN) consider other parameters that increase the accuracy of the estimations, such as: the fracture density parameter (number of fractures per length/area or volume) and fractal dimension, where the fracture center's fractality is imposed through a hierarchal multiplicative process [9]. Most of the aforementioned algorithms (and codes) impose a power law or modified power law distribution to fracture length. The cut-off value in power law distribution shows a collapsing data set, indicating a similar mechanism/signature in the modeled data set. The basic idea behind the power law distribution of fractures remains a challenging question. However, recent developments in graph theory have tried to provide a reasonable answer.

In three-dimensional scenarios, each individual fracture is assumed as triangle, circle or any other polygon shapes where the consecutive generations based on the pre-set parameters are accomplished [17]. Another main attribute of the fracture system is the density or spatial density of fractures. This property yields the possible concentration of fractures in space as well as fracture tips, around tunnels, the interface between two layers and generally any sharp or disturbed sub-fields within the system. The equivalent definition of such fracture zones may be found in modularity and the "hubness" concept, in which groups, communities or dense clusters, through random links, communicate with each other [18]. Our algorithm captures the power law distribution, directionality of the joint sets and the "hubness" properties of the networks, as described below (Figure 1).

This algorithm is based on power law distribution of fracture length $p(l) \sim l^{-\gamma}$ where $l$ is the fracture length and gamma is the power that controls the connectivity and length. When gamma increases, the probability of finding fractures with long length reduces. Other parameters are: the number of joint sets and fracture zones (hubs), hub growth, background rock joint growth (hereafter called the external links or back joints), maximum diameter of growth in $x$ and/or $y$ directions, azimuths of joint sets and mean aperture of each fracture. The parameter related to "hub" and "back" growth is reported as reciprocal values. Then large values of growth show slow propagation of links. The spatial distributions of hubs are accomplished by employing a Gaussian distribution. The growth in hub zones and complementary background fractures are based on a power law distribution of fracture length and the uniform distribution of joint sets' dips. We show that, for a wide range of gamma parameters, the distribution of links—in the transformed shape of fracture networks— obeys the power law distribution, indicating scale free networks. In other words, with the described method, we achieved modular networks with power distribution in transformed networks.

```
n_g=200; %Number of Generations
    n=300;    %Number of grids
NJS =2; %Number of Joint set

    gama=.55; %parameter related to broadness of fracture
length p(l)~l^-γ;γ=0.55
    alpha=5*n./2; %parameter to scale the power law
    hub_growth=3;% reciprocal of growth rate of fractures
around hubs
    Back_growth=2;%reciprocal of growth rate of background
fractures

N.F.Z=1:3 %number of fracture zones (hubs)

r1 = round(10 + 10.*rand(1,1));
r2 = round(10 + 10.*rand(1,1)); %radius of hubs -effective zone
of fracture zones (homogeneous cores)
```

Fig. 1. Part of the parameters employed in fracture network code-The code was compiled in M-file format/Matlab.

## 2.2. *Modern Graph Theory*

A network consists of the nodes and edges connecting them [11-12]. In order to analyze the network properties of the generated fracture networks, each fracture generation was mapped into a corresponding node in a graph space. When two fractures are intersected, two nodes are connected by a link. This procedure is illustrated in Figure 2, which enables us to investigate the networks using the tools from modern network theory discussed in the following.

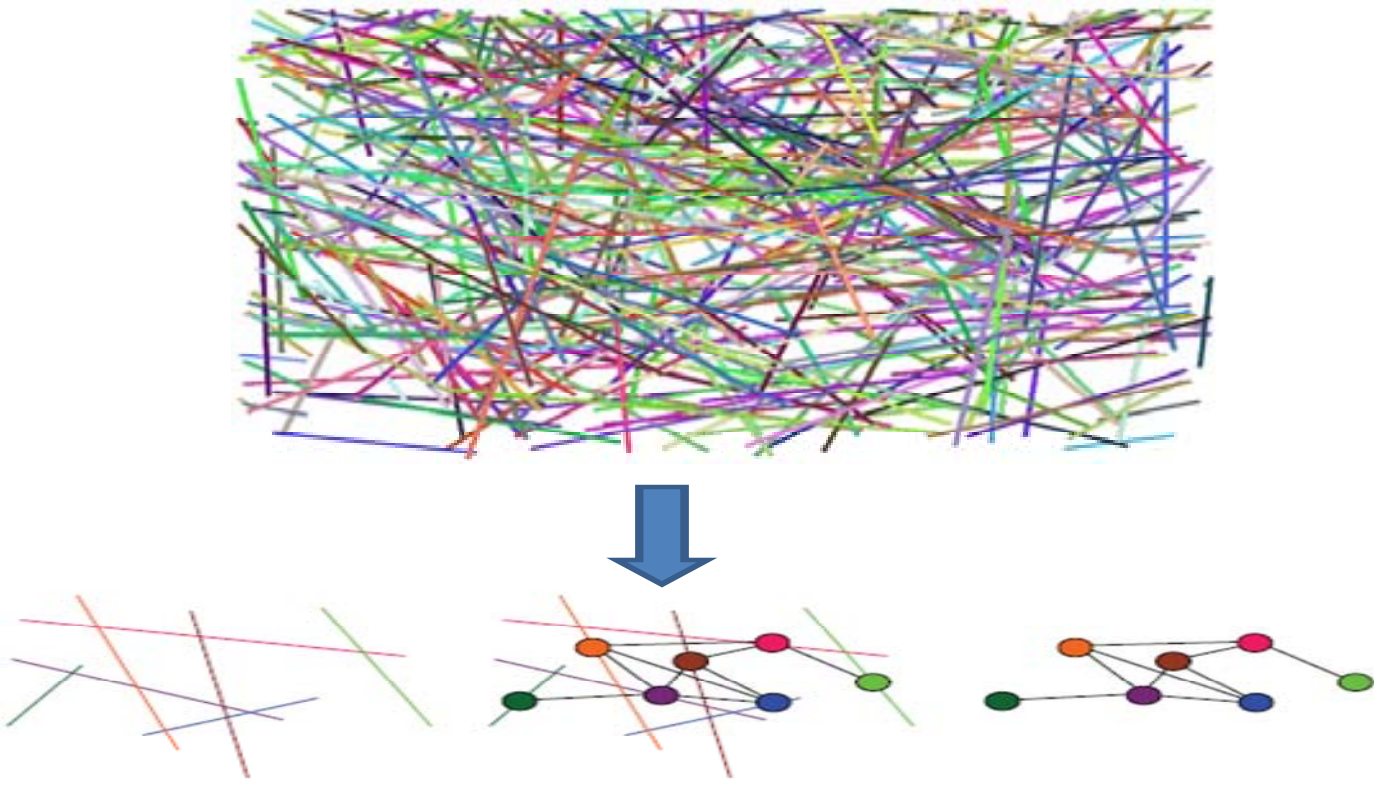

Fig. 2. The transformation procedure of a spatial fracture network in a graph shape with nodes and edges (also see [16]).

Let us introduce some properties of the networks, such as: the clustering coefficient ($C$), degree distribution ($P(k)$) and average path length ($L$). The clustering coefficient describes the degree to which $k$ neighbors of a particular node are connected to each other. Neighbors are defined as the connected nodes to a particular node. The clustering coefficient shows the collaboration between the connected nodes, i.e., the local structures. Assuming that the $i^{th}$ node has $k_i$ neighboring nodes, there can exist at most $k_i(k_i-1)/2$ edges between the neighbors. $C_i$ is defined as the ratio:

$$c_i = \frac{\text{number of edges between the neighbors of the } i^{th} \text{ node}}{k_i(k_i-1)/2} \qquad (1)$$

Then, the clustering coefficient is given by the average of $C_i$ over all the nodes in the network [12]:

$$C = \frac{1}{N}\sum_{i=1}^{N} c_i. \qquad (2)$$

For $k_i \le 1$ we define $C \equiv 0$. The closer $C$ is to one, the larger the interconnectedness of the network. The connectivity distribution (or degree distribution), $P(k)$, is the probability of finding nodes with $k$ edges in a network. In large networks, there will always be some fluctuation in the degree of distribution. Large fluctuations from the average value ($<k>$) refer to the highly heterogeneous networks, while homogeneous networks display low fluctuations [12-13]. The average (characteristic) path length, $L$, is the mean length of the shortest paths connecting any two nodes on the graph. The shortest path between a pair $(i, j)$ of nodes in a network can be assumed as their geodesic distance, $g_{ij}$, with a mean geodesic distance $L$ given as below [12]:

$$L = \frac{2}{N(N-1)}\sum_{i<j} g_{ij}, \qquad (3)$$

where $g_{ij}$ is the geodesic distance (shortest distance) between node $i$ and $j$, and $N$ is the number of nodes. Based on the aforementioned network characteristics, two lower and upper bounds of networks can be recognized: regular networks and random networks (or Erdős–Rényi networks [12]). Regular networks have a high clustering coefficient (C ≈ 3/4) and a long average path length. Random networks (whose construction is based on the random connection of nodes) have a low clustering coefficient and the shortest possible average path length. However, Watts and Strogatz [11-12] introduced a new type of network with high clustering coefficient and small (much smaller than the regular ones) average path length.

Analysis of the internal network structures, using the obtained networks, is presented by sub-graphs and motifs. The subgraphs are the nodes within the network with the special shape(s) of connectivity. The relative abundance of subgraphs has been shown to be an index to the functionality of networks with respect to information processing. They also correlate with the global characteristics of the networks [19-20]. The network motifs introduced by Milo et al. [19] are the particular sub-graphs representing patterns of local interconnections between the nodes in the network. A motif is a sub-graph that appears more than a certain amount (other criteria can be found in the literature). A motif of size $k$ (containing $k$ nodes) is called a $k$-motif (or, more generally, a sub-graph).

### 3.3. *Advection Based Network Equation*

For a given network with $N$ nodes, the degree of the node and the Laplacian of the connectivity matrix are defined by [25]:

$$k_i = \sum_{j=1}^{N} A_{ij}; L_{ij} = A_{ij} - k_i \delta_{ij} \qquad (4)$$

where $k$, $A_{ij}$, $L_{ij}$ are the degree of $i^{th}$ node, elements of a symmetric adjacency matrix, and the network Laplacian matrix, respectively. The flow of information ($u_i$) is expressed by the network's mode of diffusion or advection [26]:

$$\frac{d}{dt} u_i(t) = f(u_i) + \varepsilon \sum_{j=1}^{N} L_{ij} u_j \qquad (5)$$

where $f(u_i)$ and $\varepsilon$ are the local dynamics of each profile (node) and diffusion constant, respectively (in our simulation $f(u_i) = 0; \varepsilon = 1$). This equation is a discrete form of the classical diffusion (advection) equation. In a steady state case, we have $\varepsilon \sum_{j=1}^{N} L_{ij} u_j = 0$ which is equivalent with $\nabla^2 u = 0$ while each node is connected to $k$ of other nodes instead of 2 or 4 nodes. The application of Eq. 12 as coupled oscillators also has been investigated in synchronization (or the reaching of a steady time state) of oscillators mounted on each node.

## 3. LATTICE BOLTZMANN METHOD (LBM)

Historically, the LBM results from efforts to improve the Lattice Gas Cellular Automata (LGCA) [22]. The LBM predicts the distribution function of fictitious fluid particles on fixed lattice sites for discrete time steps in discrete directions. Two main steps in the LBM algorithm are the streaming and collision. In the streaming step, particles move to the nearest neighbor along their direction-wise discretized velocities. The streaming process is followed by the collision step, where the particles relax towards a local equilibrium distribution function. In the single-relaxation time LBM, the collision operator is simplified to include only one relaxation time for all the modes. In the D2Q9 LBM, the fluid particles at each node are allowed to move to their eight nearest neighbors with eight different velocities, $\mathbf{e_i}$. The ninth particle is at rest and does not move. The fluid density and macroscopic velocities are calculated by properly integrating the particle distribution functions on each node. The evolution of distribution function in single-relaxation time LBM [22-24] is given by:

$$f_i(\mathbf{x} + \mathbf{e_i}\Delta t, t + \Delta t) = f_i(\mathbf{x}, t) - \frac{\Delta t}{\tau}[f_i(\mathbf{x}, t) - f^{eq}{}_i(\mathbf{x}, t)] \qquad (6)$$

where for any lattice node, $x + e_i \Delta t$ is its nearest node along the direction $i$; $\tau$ is the relaxation time; and $f^{eq}{}_i$ is the equilibrium distribution function in direction $i$. The macroscopic fluid variables, density $\rho$ and velocity can be obtained from the moments of the distribution functions as follows:

$$\rho = \sum f_i; \rho\mathbf{v} = \sum f_i \mathbf{e_i} \qquad (7)$$

while the fluid pressure field $p$ is determined by the equation of the state of an ideal gas:

$$p = C_s{}^2 \rho \qquad (8)$$

where $C_s$ is the fluid speed of sound and, for D2Q9, the lattice is equal to $\frac{1}{\sqrt{3}}$. The relaxation time characterizes the time-scale behavior of fluid particle collisions and determines the lattice fluid viscosity:

$$\nu = \frac{1}{3}(\tau - \frac{1}{2}) \qquad (9)$$

Once the macroscopic velocity field is determined, the permeability of the medium under study can be predicted using Darcy's law:

$$\langle \mathbf{v} \rangle = -\frac{\mathbf{K}}{\mu} . \nabla p \qquad (10)$$

where $\langle \mathbf{v} \rangle$, $\kappa$, $\nabla p$ and $\mu$ are the volume averaged flow velocity, permeability tensor, pressure gradient vector and the dynamic viscosity of the fluid, respectively. The fluid flow is driven by pressure (or by some external force) on the boundaries. To extract the correct permeability values, the system should be in the steady state. The computation is in the steady state when the relative change of the average velocity is less than a tolerance variable (here $10^{-9}$). More details about the different numerical and technical aspects of the LBM can be found elsewhere [22-24].

## 4. RESULTS AND DISCUSSION

In this section, the results of our simulation, based on the aforementioned methods, are presented. Two-dimensional fracture networks were generated at the percolation threshold, where there is a continuous path from one side of the system to the other. All of the cases had 270 *270 grids except one case which includes 650*650 grids. The first case (upper row in Figure 3a) was a complex one that included a uniform distribution of apertures for the fractures examined. However, in all of the other cases, similar values of opening (aspect ratio) were chosen for the generated fractures. As it has been depicted in Figures 3 and 4, increasing the gamma parameter, fracture length and increases the number of fractures with small length. This changes the total permeability and velocity ditribution of fluid. Furthermore, the increment of gamma parameter induces a less complex fluid pattern (Figure 4). A broad range of directionality in fracture generations also dramatically changes the complexity of the flow. This issue is revealed in the curvature of the flow paths, i.e. the tortuosity. It induces more consumption of energy in order to reach steady state or the flow to drive. The fracture systems with regular rock joint sets[1] exhibit relatively simpler, and therefore more predictable, fluid flow paths (Figure 5). Particularly, the relative direction of flow to the dip of fracture sets changes the spread of permeability (the range of variation of permeability [Figure 13]). The distribution of velocity directly affects the total permeability of the system as well as synchronization time (time to reach steady state)[2].

The results of the modeling with LBM and FEM (Figures 4, 5 and 7) show distinguished abnormality and differentiated spatial area in velocity distribution. As it has been visualized in figures 4 and 5, the dense fractures resulted from low values of gamma parameters having more complicated discretized areas (blocks) than regular rock joint sets. We detect 4-edged blocks (as loops) and other non-loops 4-point nodes in transformed graphs[3]. It has been shown that the general configurations of the generated systems vary, but that the system ensures a nearly identical distribution of 4-point loops/non-loops. This is correct for a broad range of parameter variations in terms of gamma. In the figure above, gamma is spelled as 'gama'. Should it be gamma there too?

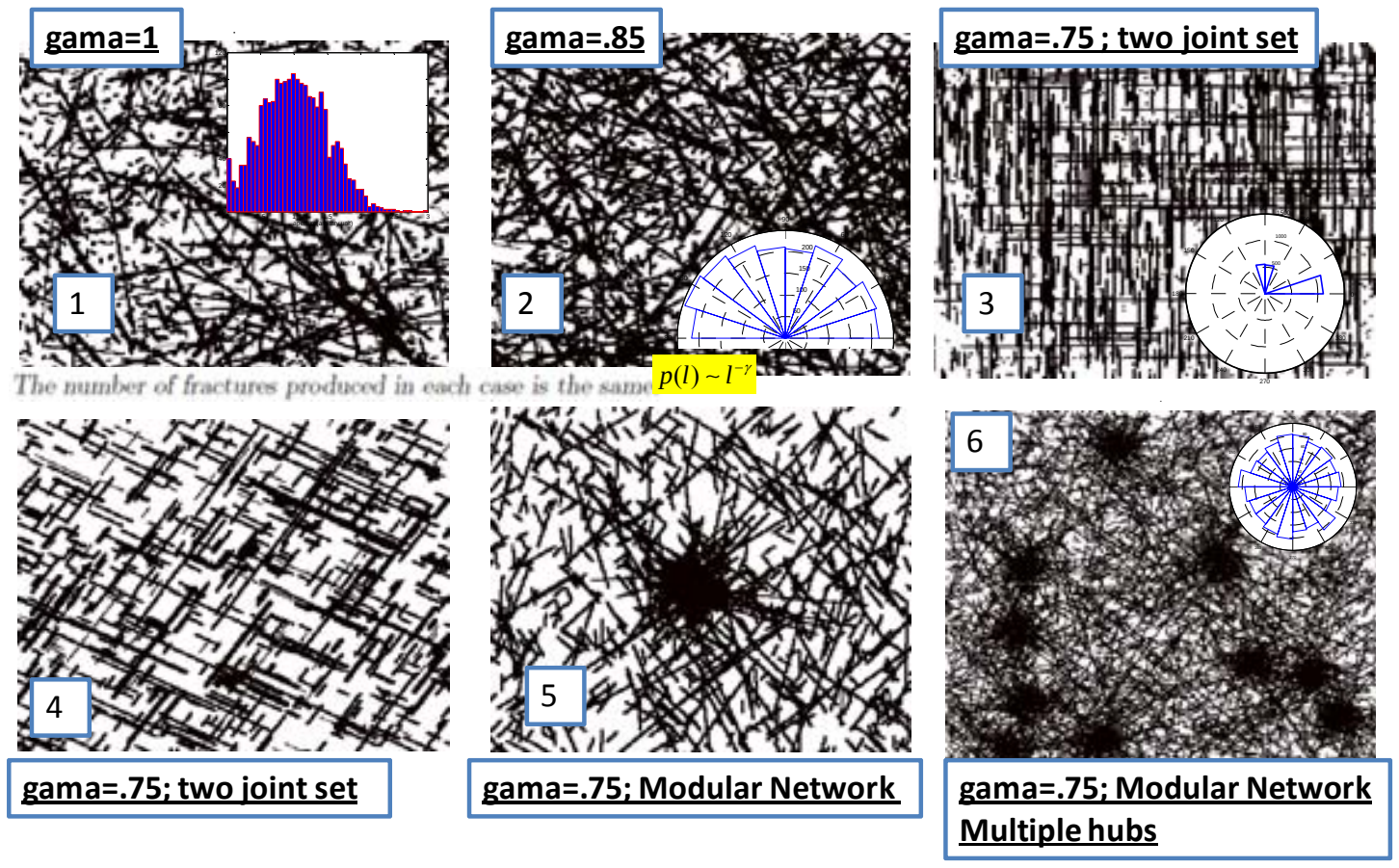

Fig. 3.Variation of fracture networks depends on "gamma" parameter joint sets and the density of fractures. The angular distribution of joints and fracture length has been shown as the insets.

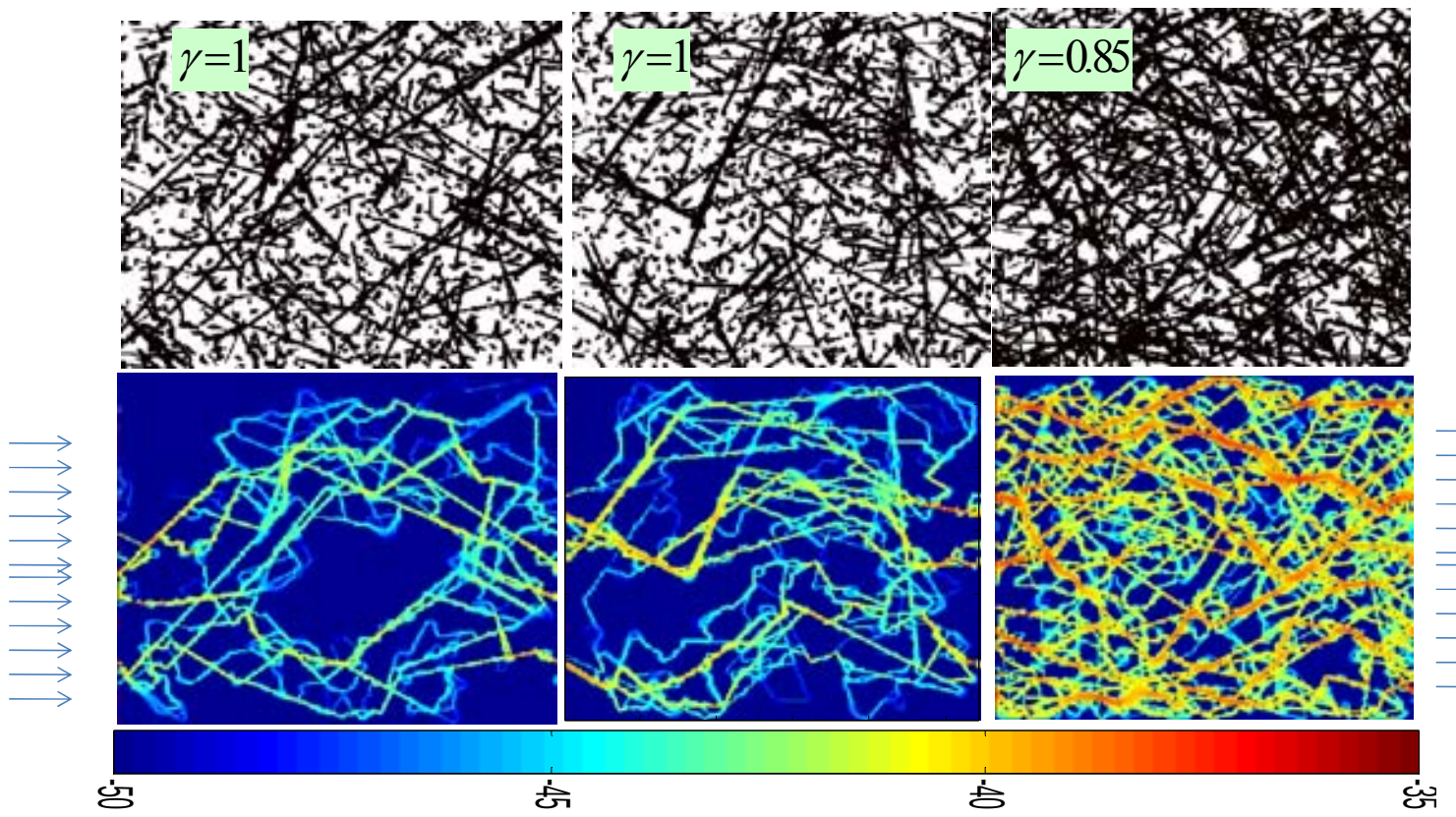

Fig. 4. Complex fracture networks with different distribution in fracture length and corresponding fluid flow patterns, obtained by LBM. The velocity field is in logarithmic scale.

To model the heterogeneity in fracture density, we set a uniform core for each module (fracture zone) with a maximum 10 or 20 units (arbitrary) radius, and with star-like directionality (with uniform distribution). Background random links have a uniform distribution of dip in the range of 0-180 (half star). Modeling with LBM (and also with FEM) shows how fracture zones trap the flow and change pressure distributions (Figure 6 and 7). Generally, the fracture zones act as the relaxation of velocity and rapid reduction of velocity (and then the gradient of pressure). More fracture zones generally reduce the permeability of the total system; however, its reduction magnitude depends on quality of links among modules (background rock joint).

---

1 Hereafter, "regular rock joints" stands for joints with sets (two or three).

2 For example see: Permeability of Three-D Random Fiber Webs by Koponen et al in PRL-VOL.80-1998

3 It can be investigated further that transformed networks produce sub-graphs with different shapes.

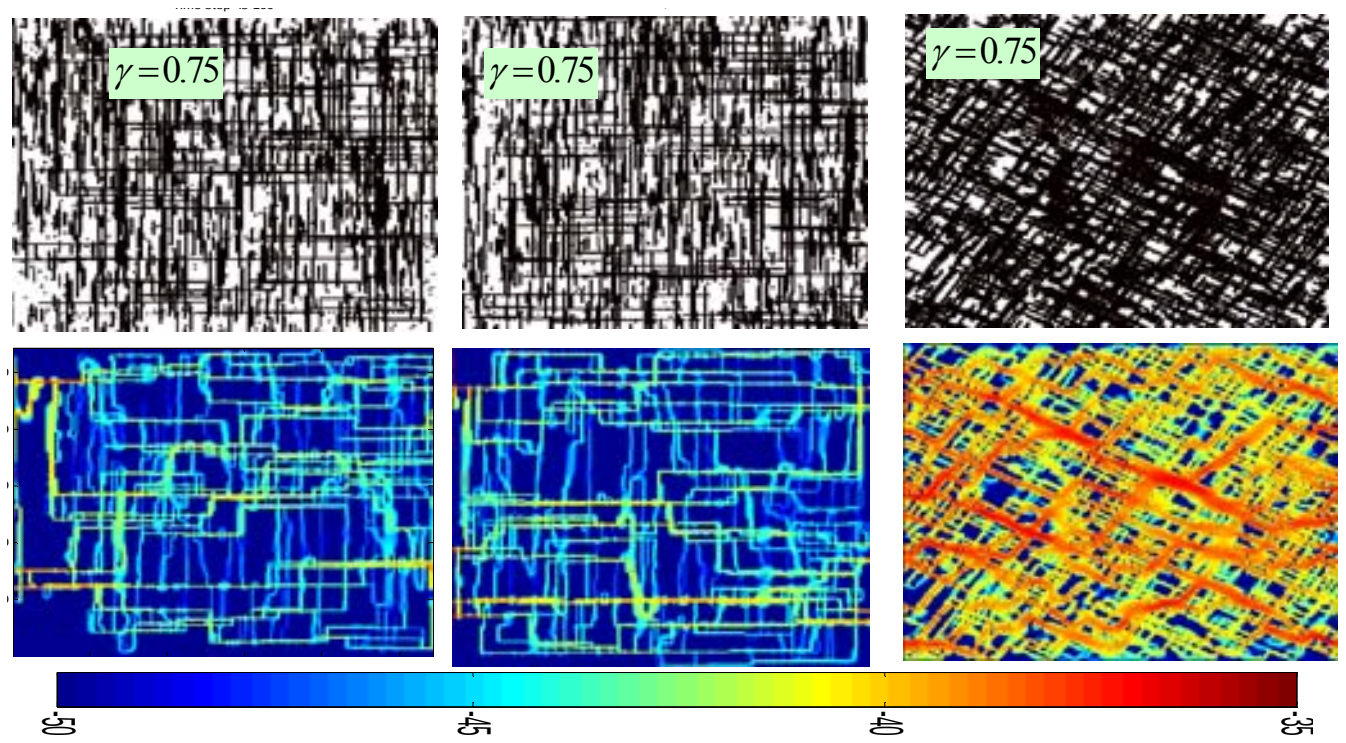

Fig. 5. Other configurations of complex fracture networks with different distributions in fracture length, the directionality of joint sets with two sets and corresponding fluid flow patterns, obtained by LBM. The velocity field is in logarithmic scale.

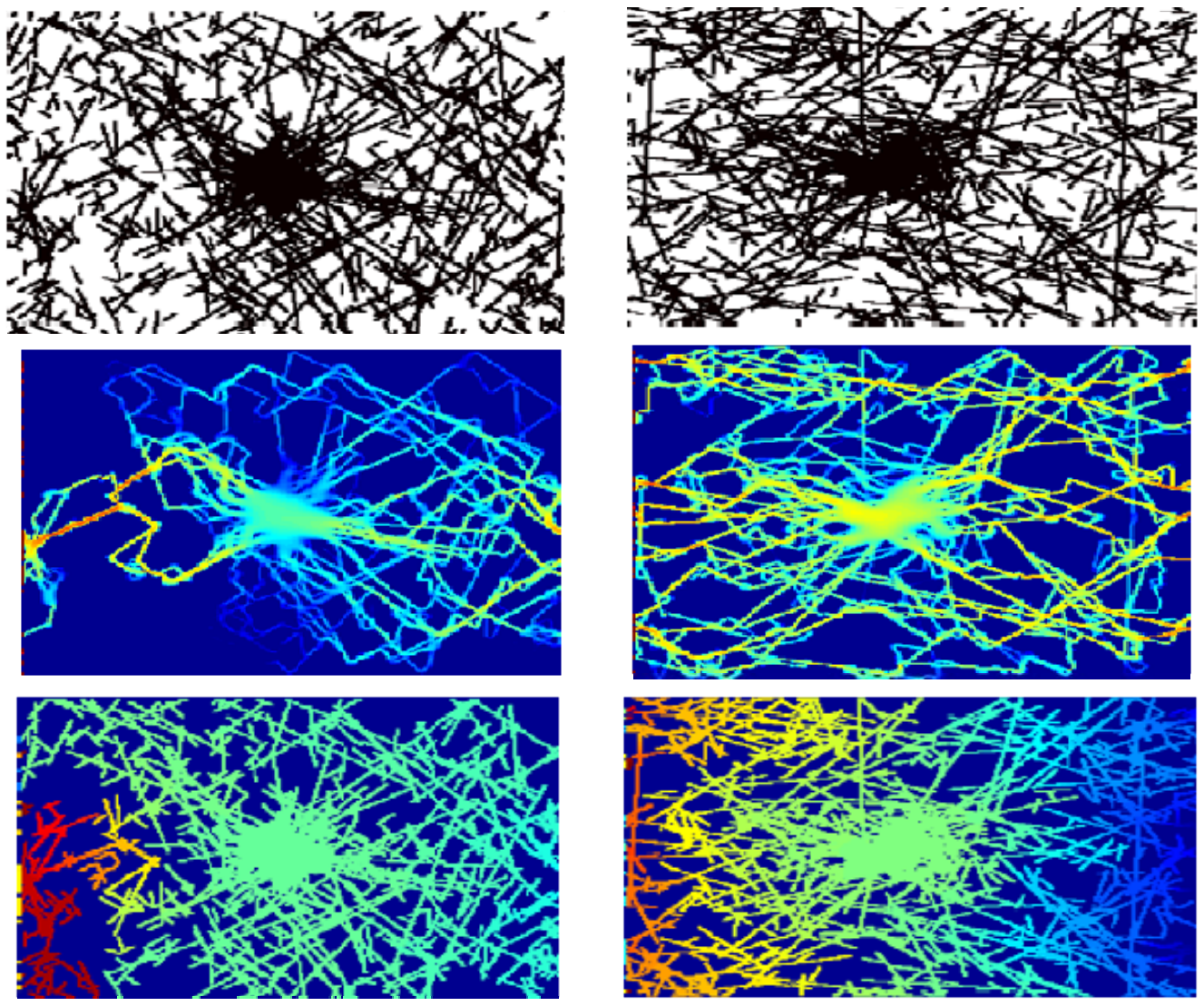

Fig. 6. Left) Single hub pattern with gamma =0.75, hub growth=8, back growth =3, its velocity distribution using LBM and pressure distribution; Right) Single hub pattern with gamma =0.75, hub growth=10, back growth =3, its velocity distribution and pressure distribution, from up to down.

Two general points should be considered in the analysis of permeability: the influence of internal links in the fracture zones and the effects of external links. Each of these parameters effectively changes the information (fluid) of the flow. Weak external links yield concentration of particles in especial modules while frequent external links reduce concentration of particles and the flow from modules. The internal fractures of the fracture zone impose local complexity, and can be assumed to be sub-fracture networks within the system. The combination of the aforementioned factors gives rise to a complicated situation

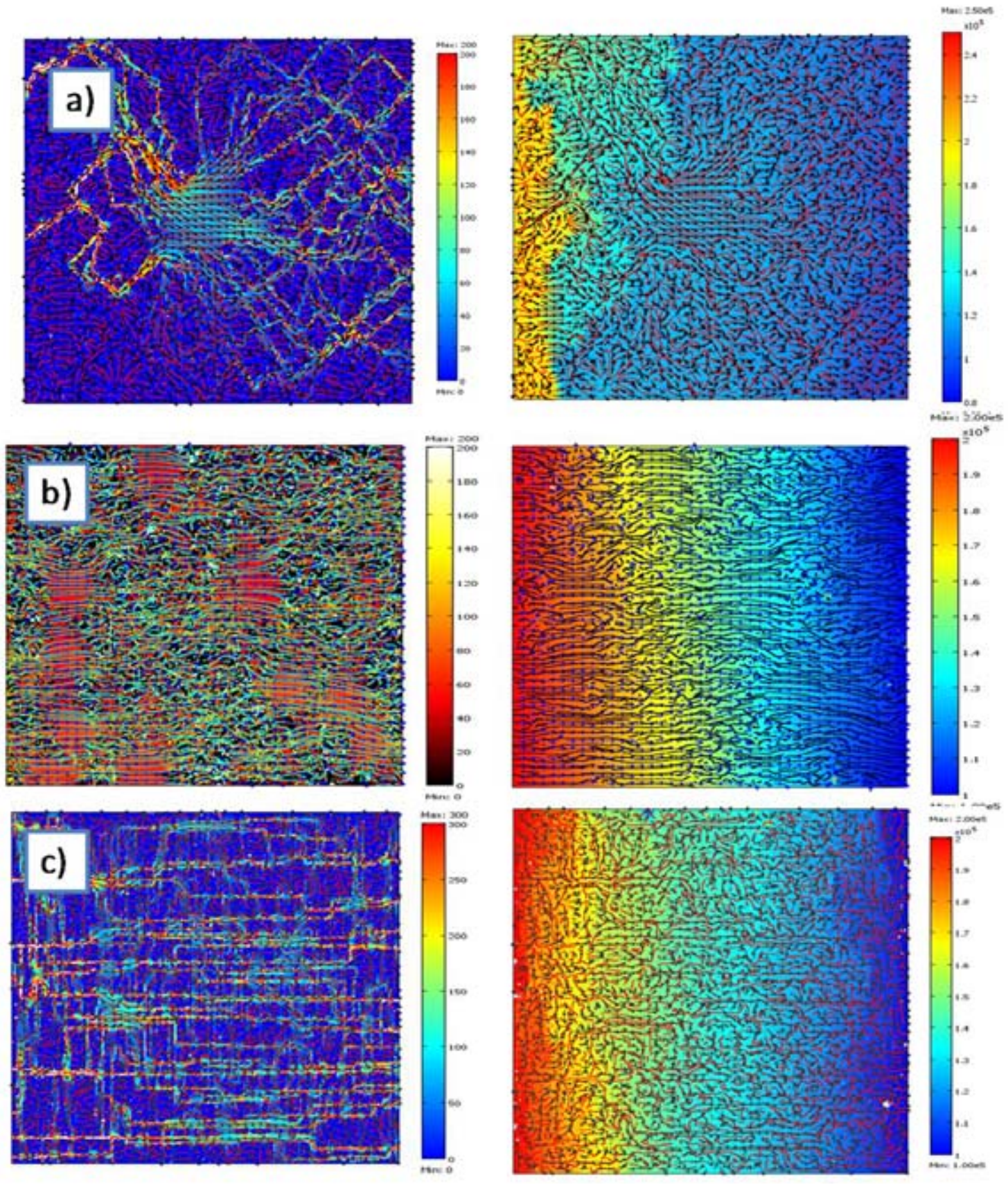

Fig. 7. Fluid flow modeling with FEM in a) Single hub pattern with gamma =0.75, hub growth=8, back growth =3; b) Multiple hubs with 650 grids and c) two joint sets (0, 90 degree azimuths). The right hand pictures are the pressure distribution with the corresponding networks.

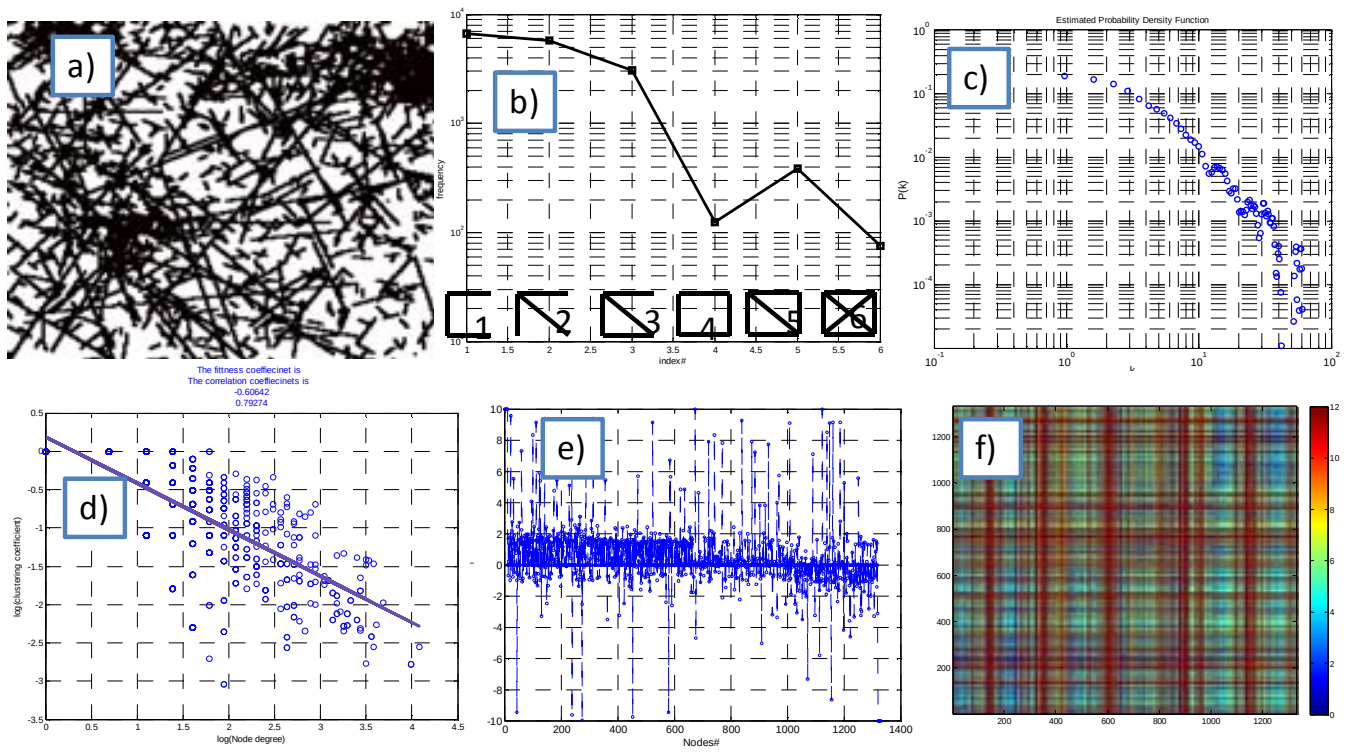

Fig. 8. a) Network with 400 generation per each hub with 4 hubs. Hub growth and back-growth parameters respectively are 3 and 2 with gamma=0.85; b) sub-graphs distribution of 4-points sub-graphs over transformed fracture networks in graphs; c) distribution of node's degree; d) scaling of clustering coefficient with number of links; e) distribution of velocity variation in advection–network system; and f) visualization of node's distance.

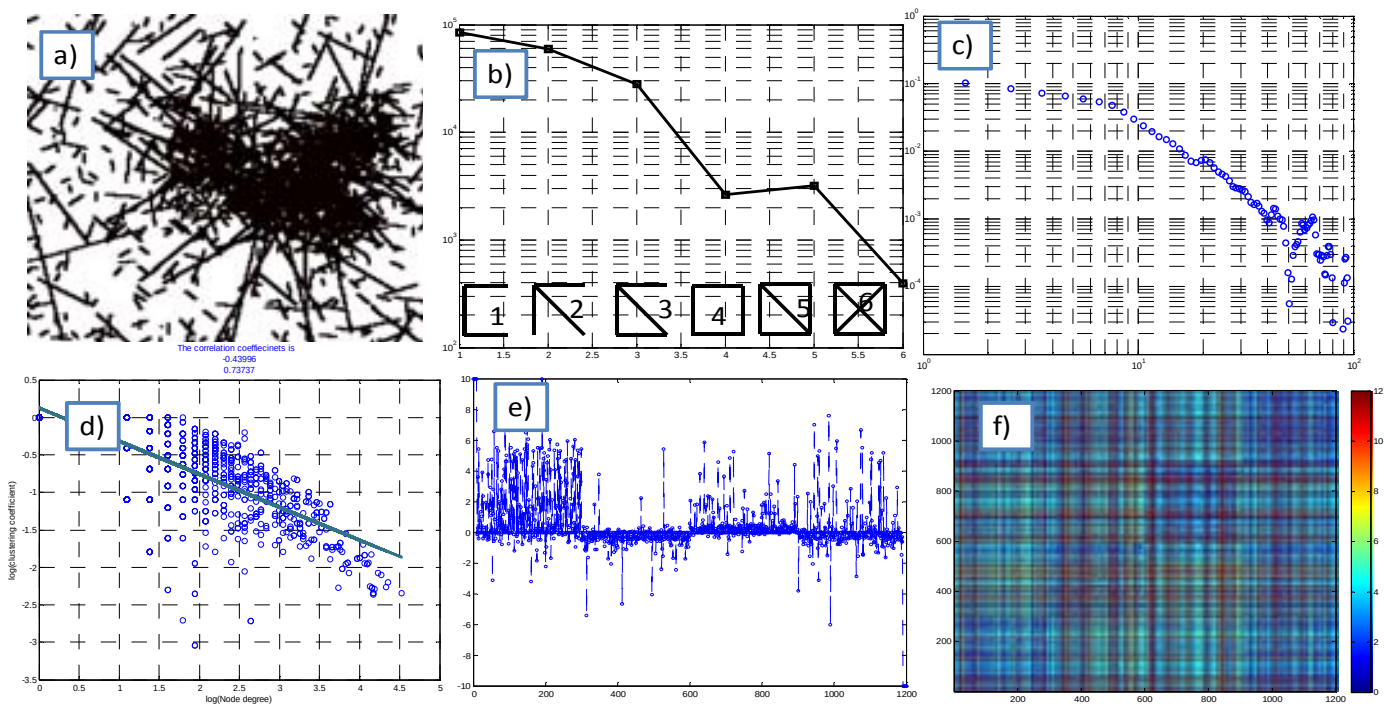

Fig. 9. a) Network with 400 generations per each hub with 4 hubs. Hub growth and back-growth parameters respectively are 2 and 4 with gamma=0.85; b) sub-graphs distribution of 4-points sub-graphs over transformed fracture networks in graphs; c) distribution of node's degree; d) scaling of clustering coefficient with number of links; e) distribution of velocity variation in advection–network system; and f) visualization of node's distance.

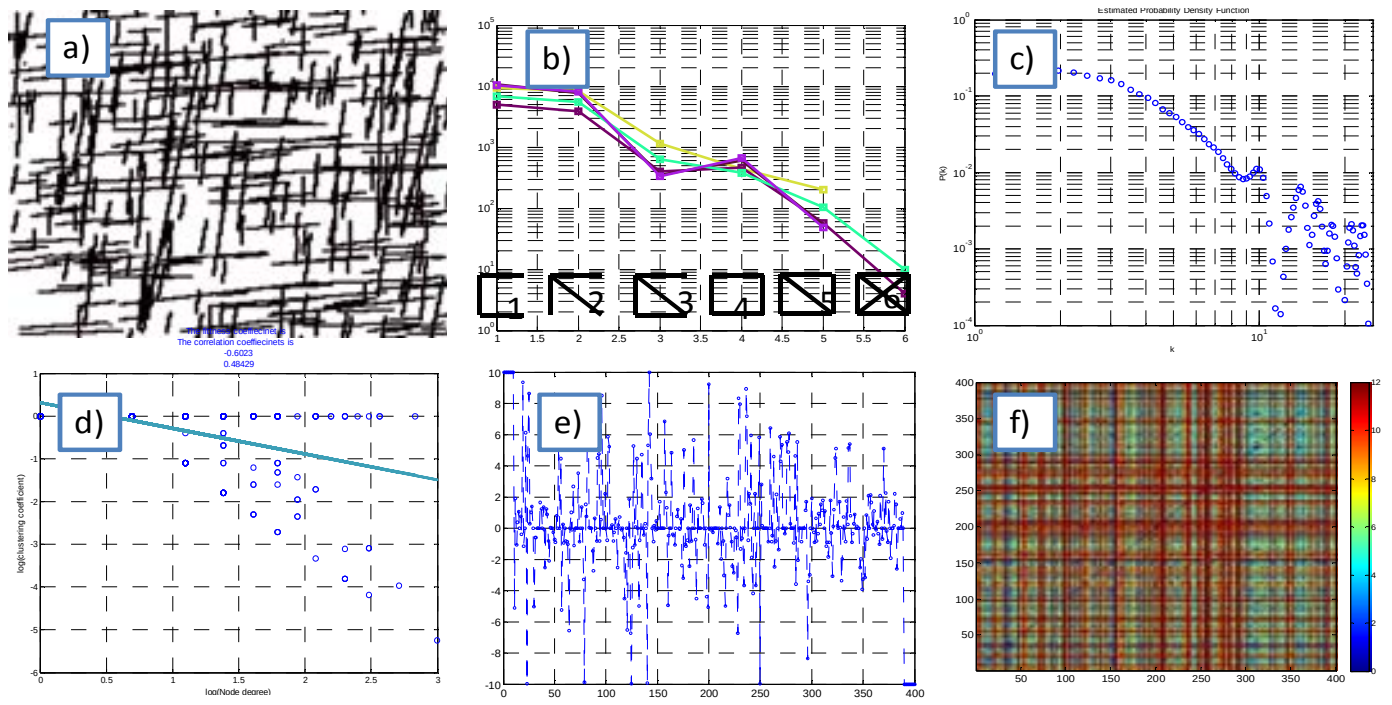

Fig. 10. a) Network with 400 generation per each hub with 4 hubs And gamma=0.85 with two rock joint sets; b) sub-graphs distribution of 4-points sub-graphs over transformed fracture networks in graphs; c) distribution of node's degree; d) scaling of clustering coefficient with number of links; e) distribution of velocity distribution in advection –network system; and f) visualization of node's distance.

It is noticed that, due to the random nature of the fractures' webs, Monte Carlo simulation is necessary to get a precise answer. For LBM, only 5-10 fractures for each case (each constant parameter in the model) have been analyzed since the procedure is time-consuming. For modeling in FEM, it has been assumed that each fracture has the dimension in z direction as the aperture /opening of each joint. The results of simulation with LBM and FEM (5, 6 and 7) show approximately the same patterns in fluid flow path. It is expected that the results for permeability or twistability will show the same answer using both methods. Now, we map the spatial joint networks into graphs, using the method described in section 2.2. It has been distinguished how different hubness parameters affect the hierarchical patterns in fractures. Using the method described in section 3.3, the distribution of velocity over nodes in steady state is presented and compared with the results obtained with the FEM method. To model Eq. 12, the Euler method has been used in integration, considering the stability of the solution with respect to $\Delta t$ and $\Delta x$ ($\varepsilon$ is constant per each fracture which is equivalent with permeability of each joint).

In figures 8 to 10, five main statistical characteristics of graphs have been shown. Part "b" in the aforementioned figures shows the distribution of 4-point sub graphs, indicated by indexes 1 to 6. Despite different configurations in hubness implementation (figures 8 and 9), the frequency of sub-graphs follows a similar trend. Decrease in the frequency of the index 4, as a rectangular loop, and increase in the index 5 are common features of the frequencies. This is not observed in regular joints with perpendicular intersection (Figure 10) in which the index 4 shows a slight increase. Referring to simulation with LBM (or FEM), it reveals that the regular fracture networks with pre-set joint sets present much more uniform distribution in the velocity than in fracture zones (all see Figure 11 with advection-network equation).

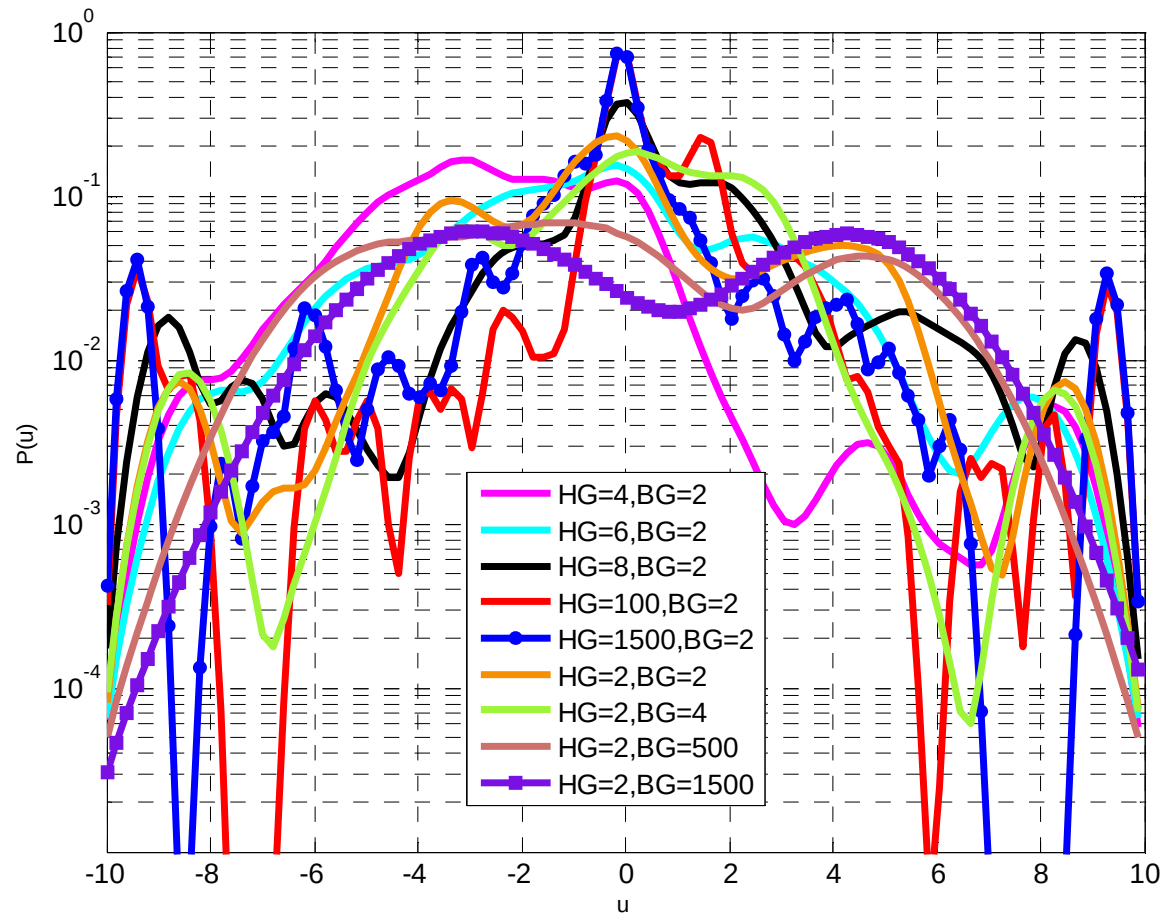

Figure 11. Distribution of velocity in transformed networks using the steady state solution of the advection-network system over different configurations of hub growth (HG) and back fracture growth (BG). The values of growths are reciprocal to the propagations of fracture. Total number (max.) of hubs could be 4. The simulated system had sinks and sources nodes with 10 and 10 values of velocity (or concentration of particles). The first and last 10 fracture networks were considered as sources and sinks respectively[4].

It suggests that the index 4 (and somehow 2[5]) implies the most channelized flow, possibly with a low value of twistability. Indexes 3 and 5 indicate such abnormality in fluid flow. Index 6 possibly shows the most

4 The results presented here are the mean results for 5 times realizations. The injection of particles from not-similar points like figures 4-7 also might be questionable, but it can be proved that the general distribution would not be affected.

5 We analyzed this fact in a single shear fracture; see H.Ghaffari et al., Fluid flow analysis in a rough fracture (type II) using complex networks and lattice Boltzmann method, PanAm-CGS (2011), Toronto, Canada.

homogeneity in information flow. Such abnormality in the distribution of velocity 6 (or particle concentration) through nodes (fractures) has been modeled with the advection based network equation described in section 3.3, and illustrated in part "e" of the figures 8 to 10. In our model, the source and sink terms were kept in the first and last 10 generations of fractures (with values with the amount 10 and -10, respectively). Increasing external links (background joints) induces informality and homogeneity in the pressure and velocity of spatial spread. In other words, it somehow suppresses the "hubness" and modularity attribute of the system. Increasing the "hubness" aspect and decreasing external links yields non-uniform flow and concentration of particles in special nodes.

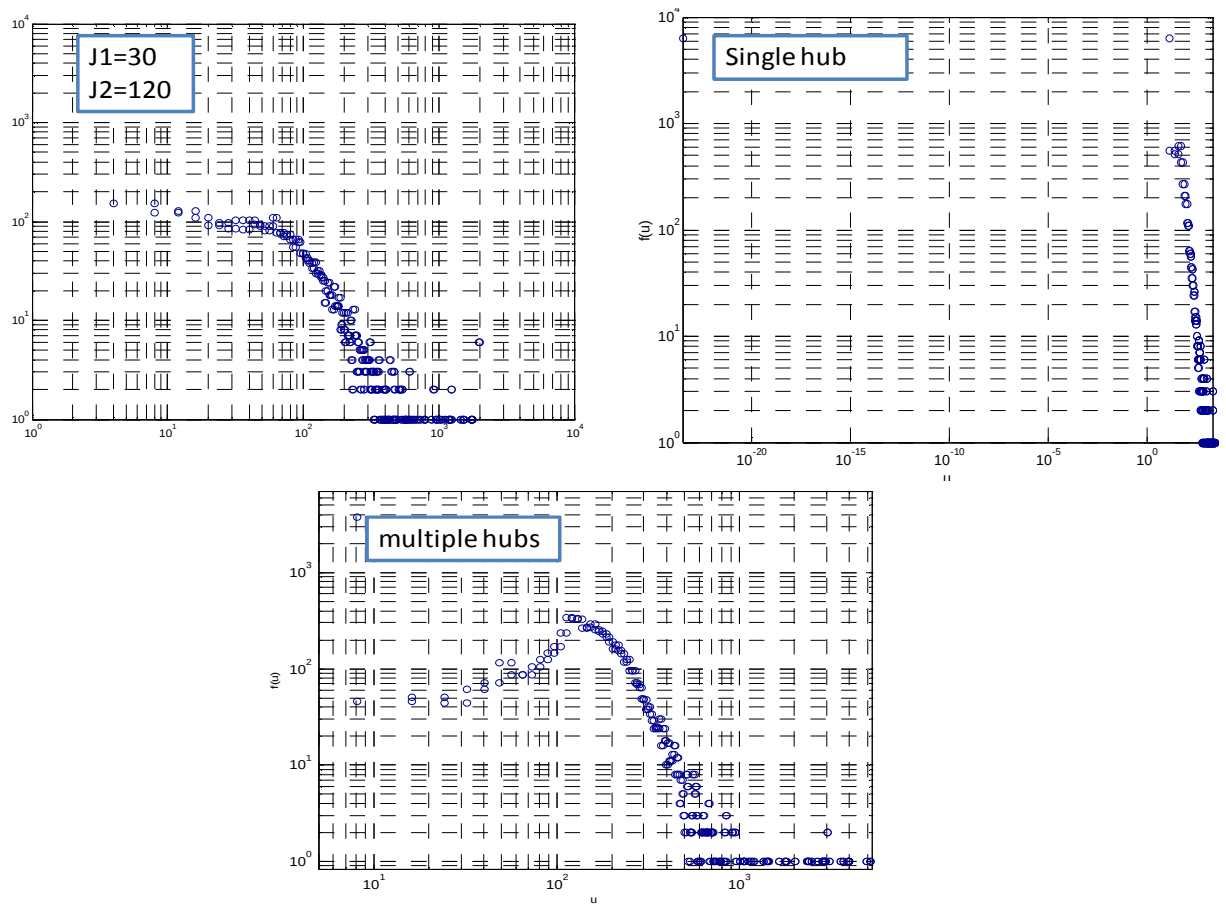

Fig. 12. Distribution of velocity in spatial fracture networks using FEM over 3 configurations of spatial fracture networks. The simulated systems have been shown in Figure 7.

Figure 11 shows the distribution of velocity fields in the steady state case for different configurations of fracture zones' parameters. The general trend of distribution is Gaussian. For the system with minimum effect of modularity, we observed more fluctuation in distribution with an evident peak around zero velocity. This indicates that decreasing modularity induces a transition in velocity (and then total permeability) distribution. Deleting or cutting external linkages forces the system to have power law distribution— i.e. abnormality in pressure and velocity abundances— with less fluctuation in velocity or pressure frequencies. Plotting the distribution of velocity field -modeled with FEM- shows a distribution with the long tail characteristic. (Figure 12). In the transformation of spatial networks, the spatiality has been lost and the same properties of the flow with such a mapping into the graphs have been recovered. Part "c" in the above figures shows the degree distribution, which roughly obeys power law distribution. In other words, our fracture algorithm (with or without modularity), for a wide range of parameter variations, gives "scale-free networks". Part "d" demonstrates a space called the spectrum of graphs, which is associated with the clustering coefficient ($c$) - node's degree ($k$) space. It has been shown that networks with power law distribution with a linear scaling of $c$-$k$ present with a hierarchical property [27]. Strong hierarchical patterns in modular networks and weak hierarchical patterns in regular fracture systems have been observed (Figure 10).

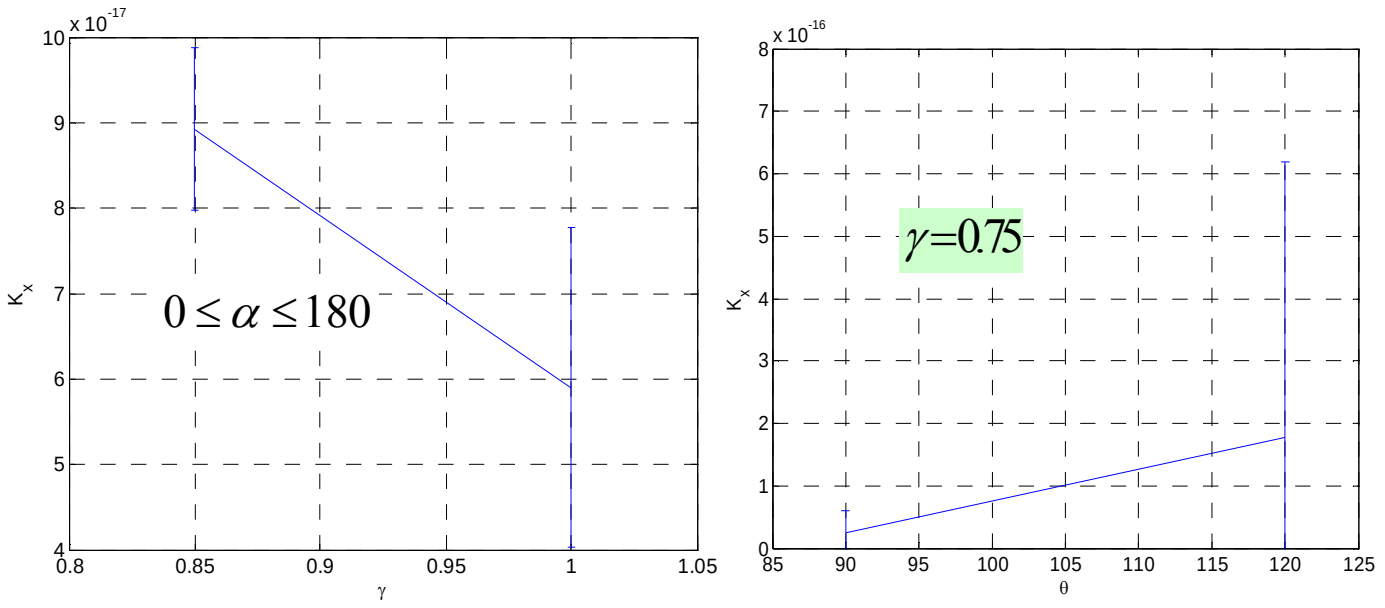

Fig. 13. Left) Variations of permeability with gamma parameter through complex fracture network configuration; Right) variations of permeability with the directionality (azimuth) of joint sets under constant gamma parameter (no hubness property is induced in facture patterns).

As the last part of our research, we focus on the possible relation between the mean path length of transformed networks and variation of permeability. Part "f" in figures 8-10 shows the distance of nodes, where cold colors indicate near neighborhoods or immediate intersected joints and hot colors indicate far neighborhoods or non-connected nodes. Comparing the calculated mean path length or the graph's diameter and the simulated permeability shows a reverse correlation between the two parameters. Rich neighborhood attribute exposes a highly permeable fractured system, while poor nodes reveal low permeability from a connectivity point of view (Figure 14).

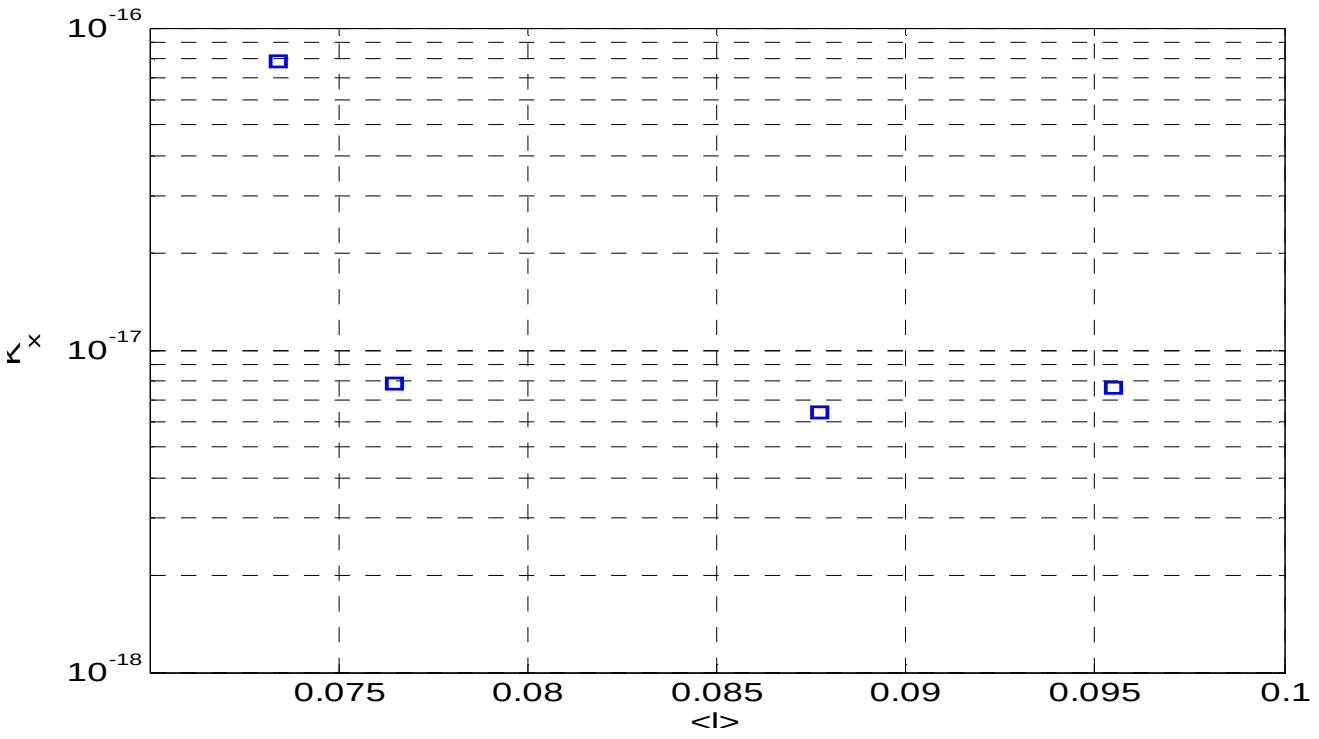

Fig. 14. An approximate relation between the mean path lengths of transformed fracture networks and simulated permeability using LBM. Each point results from 5 time realizations.

---

6 Recent analysis in network theory has highlighted the abnormality of diffusion with random walk theory and stochastic particle simulation, for example see PRL 94,248701(2005) and PRE E82,055101(2010).

## 5. CONCLUSIONS

Fracture networks in rock masses have been studied in this paper. The fluid flow in fracture networks, as a function of variation of connectivity patterns, was analyzed. The Lattice Boltzmann method and FEM were used in modeling permeability and fluid velocity distribution. Furthermore, fracture networks were mapped into graphs, and the characteristics of the obtained graphs were compared with the main spatial fracture networks. The "hubness" character of fractures as fracture zones has been distinguished. The results showed that fracture zones, the internal and external links of the damaged zones, dramatically change the transport properties of a rock mass.